\documentclass[preprint,showpacs]{revtex4}
\usepackage[dvips]{graphicx}
\usepackage{epsfig}
\usepackage{psfrag}
\usepackage{amsmath}
\usepackage{amsfonts}
\usepackage{amssymb}
\usepackage{bm}
\begin{document}

\title{Optical absorption spectra and monomer interaction in polymers. Investigation of exciton coupling in DNA hairpins}
\author{A. L. Burin, J. A. Dickman, D. B. Uskov, C. F. F. Hebbard}  
\affiliation{Department of Chemistry, Tulane University, New Orleans LA, 70118}
\author{G. C. Schatz}
\affiliation{Department of Chemistry, Northwestern University, Evanston, IL, 60208}
\date{\today}
\begin{abstract}
We investigate the effect of exciton coupling on the optical absorption spectrum of polymer molecules under conditions of strong inhomogeneous broadening. We demonstrate that the dependence of the maximum in the rescaled absorption spectrum on the number of monomers is determined by the average monomer excitation energies and their resonant coupling and insensitive to the inhomogeneous broadening. Thus the absorption spectrum can be used to determine optical interactions between monomers. The results are applied to the absorption spectra of poly-A poly-T DNA hairpins and used to interpret the dependence of the absorption spectrum on the number of monomers. We also discuss exciton localization in these hairpins.  
\end{abstract}


\maketitle

{\bf 1.} 
The optical absorption spectra of dissolved molecules are strongly affected by interaction with the environment. The positions and orientations of solvent molecules with respect to each molecule differ, and leads to different shifts in their optical excitation energies. These shifts can be characterized by a distribution with characteristic width $W$ that is usually called inhomogeneous broadening. In many important systems, particularly including biological molecules in their native environment involving a highly polar solvent (water), this broadening is very large $W>0.1$eV and it can smear out various interesting features of the spectrum including its sensitivity to molecular conformation and geometry.  

However there exist experimental indications that despite the large inhomogeneous broadening the absorption spectrum contains important information about molecular properties. Recent measurements of optical absorption spectra of DNA hairpins made of identical $AT$ base pairs demonstrate a remarkable change in absorption maxima with the number of monomers \cite{Fred}. We believe that this effect is due to sensitivity of the electronic excited state to the number of monomers. The excited state (exciton) can be delocalized between all $n$ identical monomers and the resulting exciton energy (absorption energy) is sensitive to the size of the overall wavefunction. The exciton coupling $V_{ij}$ of different monomers is caused by their resonant interactions, and it reaches its maximum value $V_{0}$ for adjacent monomers. One can therefore expect that the change in the absorption spectrum with $n$ is determined by $V_{0}$. Since the multipole interactions that determine $V_{0}$ are very sensitive to the distance between monomers and to their relative orientation, the change in absorption spectrum with $n$ can be used to study DNA conformation and structure \cite{books}. Thus time-resolved optical measurements of DNA participating in biological or chemical processes like protein attachment can provide important information about the real time kinetics of such process and DNA structural and conformational changes during the process. In addition, the understanding of exciton coupling is important for DNA applications as  a naturally available optical material.

\begin{figure}[b]
\centering
\includegraphics[width=9cm]{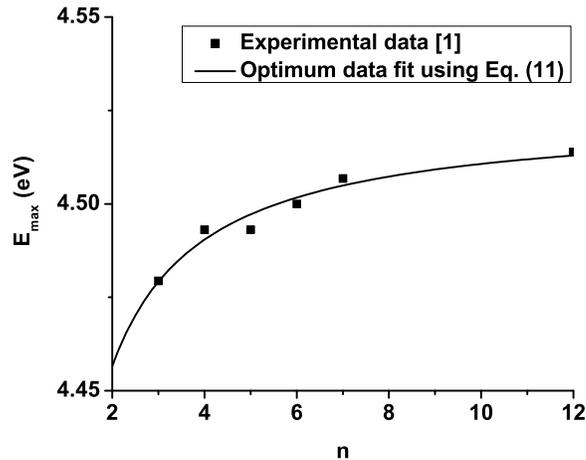}
\caption{Theoretical fit of experimental data \cite{Fred} for the dependence of maximum in the scaled absorption intensity $\beta(E)$ on the number of monomers for DNA hairpins $(AT)_{n}$.  
\label{fig:1} }
\end{figure}

Below we investigate the inhomogeneously broadened spectrum of a polymer molecule made of $n$ identical monomers assuming that inhomogeneous broadening $W$ exceeds both the exciton coupling $V_{0}$ and the variation  $\Delta$ of the average excitation energies of the different monomers with respect to each other caused by their different positions (e. g. monomers at the edges compared to monomers in the middle of the polymer). Under these conditions the absorption spectrum $\alpha(E)$ has a single maximum corresponding to electronic excitation. The position of the maximum is the important measurable characteristic of the spectrum which can be easily extracted from the experimental data. It turns out that the more informative characteristic is the maximum in the rescaled absorption $\beta(E)=\alpha(E)/E$. We investigate the dependence of the maximum in $\beta(E)$ on the number $n$ of monomers, showing that in the limit of strong inhomogeneous broadening this dependence is determined by the exciton coupling $V_{0}$ and the monomer energy variation $\Delta$, while it is insensitive to the inhomogeneous broadening. If fluctuations in the monomer energies are independent and obey Gaussian statistics then the change in the maximum position is determined by the simple analytical expression 
\begin{equation}
\delta E_{max} = \frac{4(n-1)V_{0}\cos(\theta)}{n},
\label{eq:ANS0}
\end{equation}
where $\theta$ is the angle between the transition dipole moments of adjacent monomers. As we show below this analytical result with the parameter taken to be $V_{0}\approx 0.042$eV, as obtained from semiempirical calculations and the DNA twisting angle $\theta =36^{0}$, agrees with the results of recent experiments (Ref. \cite{Fred} and Fig. \ref{fig:1}). The localization of excitons in DNA will be discussed in the light of these findings.

{\bf 2.} 
Optical excitation in a chain of $n$ monomers can be described by a tight binding Hamiltonian coupled to the environment. Differences in the environment for different monomers lead to fluctuations in the monomer excitation energies $\phi_{i}$
\begin{eqnarray}
\widehat{H}= \sum_{i\neq j}V_{ij}c_{i}^{+}c_{j}+\sum_{i=1}^{n}\phi_{i}c_{i}^{+}c_{i}. 
\label{eq:Hamiltonian}
\end{eqnarray}
Here $c_{i}$, $c_{i}^{+}$ are operators for creation and annihilation of an exciton in a site (monomer) $i$. The excitation energies $\phi_{i}$ are taken to have random values that are characterized by a distribution function $P(\phi_{1}, \phi_{2},  ..\phi_{n})$ and they are directly coupled to the local exciton population $n_{i}=c_{i}^{+}c_{i}$. All monomers $i=1, 2, ...n$  possess transition dipole moments ${\bm \mu}_{i}$ having identical absolute values $\mu_{0}$. Since the photon absorption process is very fast one can the treat energies $\phi_{i}$ as static, because their characteristic time of change is defined by a slow atomic motion.  For the same reason we only consider vertical excitations, as the environment cannot relax to the new configuration during the short absorption event. 
We ignore fluctuations of exciton couplings $V_{ij}$ due to environmental fluctuations as this interaction has a power law dependence on interatomic distances that results from multipole interactions. Since fluctuations in the interatomic distances are much smaller than the distances themselves, the power law dependent interactions should fluctuate weakly. 

The position of absorption maximum depends on the distribution function $P(\phi_{1}, \phi_{2},  ..\phi_{n})$. Below we assume that this energy distribution function can be approximated by a Gaussian distribution. This is a reasonable assumption when fluctuations of energy are smaller than the energy itself which justifies the use of a second order expansion of the free energy for the probability distribution $P$ \cite{ab2}. On the other hand the width of the distribution (inhomogeneous broadening $W \sim 0.2$eV) is much larger than the natural linewidth $\hbar \Gamma\leq 10^{-3}$eV, which allows us to ignore deviations from the Gaussian lineshape due to Lorentzian tails in the natural line. Of course calculations can be easily extended to any other distribution $P$.

If the distribution $P$ is a Gaussian distribution then all other distributions  for the reduced number of energies obtained by the integration of $P$ with respect to the remaining energies are also Gaussian distributions. For our study we will need distributions either for a single energy or for two energies. We introduce the single energy distribution as
\begin{equation}
P_{i}(\phi)= \frac{1}{\sqrt{2\pi}W_{i}}\exp\left(\frac{(\phi-\ \tilde{\phi}_{i})^{2}}{2W_{i}^{2}}\right),
\label{eq:singleE}
\end{equation}
where $\phi_{i}$ is the average excitation energy of the monomer $i$ and $W_{i}$ is its dispersion (inhomogeneous broadening). The distribution function for two energies can be expressed as 
\begin{equation}
P_{ij}(\phi, \phi')= \frac{e^{\frac{(\phi_{i}-\tilde{\phi}_{i})^{2}}{2W_{i}^{2}(1-\kappa_{ij}^{2})}+\frac{(\phi_{i}-\tilde{\phi}_{i})^{2}}{2W_{j}^{2}(1-\kappa_{ij}^{2})}-\frac{\kappa_{ij}(\phi_{i}-\tilde{\phi}_{i})(\phi_{j}-\tilde{\phi}_{j})}{W_{i}W_{j}(1-\kappa_{ij}^{2})}}}{2\pi \sqrt{1-\kappa_{ij}^{2}}W_{i}W_{j}},
\label{eq:twoE}
\end{equation}
where the parameter $\kappa_{ij}$ describes correlations of the excitations of monomers $i$ and $j$. Note that $<(\phi_{i}-\tilde{\phi}_{i})(\phi_{j}-\tilde{\phi}_{j})>=\kappa_{ij}W_{i}W_{j} $. We further assume that all average energies $\phi_{i}$ are close to some typical energy $\phi_{0}$ and their deviations from it are small compared to the inhomogeneous broadenings $W_{i}$. The same is assumed for exciton coupling so for further considerations we set 
\begin{equation}
V_{0}, |\tilde{\phi}_{i}-\phi_{0}|\ll W_{i}.
\label{eq:conditions}
\end{equation}
Here we have assumed that the energy gap to the next excited state exceeds the inhomogeneous broadening so we can neglect it.

{\bf 3}.
One can describe the optical absorption of the ensemble of polymers in terms of exciton eigenstates $\psi_{a}>, a=1, ... n$, their energies $E_{a}$, transition dipole moments $\mu_{a}$ and oscillator strengths $f_{a}$ (Eq. (\ref{eq:Hamiltonian})) as
\begin{eqnarray}
\alpha(E) \propto \sum_{a=1}^{n}<f_{a}\delta(E_{a}-E)>,  
\nonumber\\
f_{a}=\frac{2\mu_{a}^{2}mE_{a}}{3(e\hbar)^{2}},
\label{eq:AbsDef1}
\end{eqnarray}
where $e$ and $m$ are the charge and mass of electron. 

Consider the exciton coupling $V$ as a perturbation. In the zeroth order approximation in $V$ the eigenstates
$a$ coincide with individual monomer $i$ excitations characterized by energies $E_{i}=\phi_{i}$ and all
the transition dipole moments are identical, $\mu_{i}=\mu_{0}$, so we get 
\begin{eqnarray}
\frac{\alpha^{(0)}(E)}{E\mu_{0}^{2}} \propto \sum_{i=1}^{n}\int_{-\infty}^{+\infty}d\phi P_{i}(\phi)\delta(E-\phi)=\sum_{i=1}^{n}P_{i}(E).   
\label{eq:Abs0thorder}
\end{eqnarray} 
To first order in $V$, which we are interested in, there are no corrections to the exciton energy, because the off-diagonal interaction $V$ results in second order corrections only \cite{LL3}, while there is a correction to the transition dipole moment associated with the change in exciton  wavefunction $|\tilde{i}>=|i>+\sum_{j\neq i}|j> V_{ij}/(E_{j}-E_{i})$. This correction  can be expressed as $\tilde{{\bm \mu}_{i}}={\bm \mu}_{i}+\sum_{j\neq i}{\bm \mu}_{j} V_{ij}/(E_{j}-E_{i})$. Accordingly the change $\delta f_{i}$ in the oscillator strength can be expressed as $\delta f_{i}/f_{i}=2\sum_{j\neq i}\cos(\theta_{ij})V_{ij}/(E_{j}-E_{i})$, where $\theta_{ij}$ is the angle between the transition dipole moments of the monomers $i$ and $j$. Finally the rescaled absorbance can be expressed as
\begin{eqnarray}
\beta(E)=\frac{\alpha(E)}{E} \propto \sum_{i=1}^{n}P_{i}(E) +  2\sum_{i\neq j}V_{ij}\cos(\theta_{ij}) 
\nonumber\\
\times
{\rm P.V.} \int_{-\infty}^{+\infty}d\phi \frac{P_{ij}(\phi, E)}{E-\phi}.   
\label{eq:AbsRes1}
\end{eqnarray}

The factor $E$ in the definition of $\alpha(E)$ makes the absorption maximum sensitive to inhomogeneous broadening. To study lineshape effects it is more convenient to use the maximum in the rescaled absorbance $\beta(E)=\alpha(E)/E$, which represents the spectral density. Consider the energy corresponding to the  maximum in $\beta(E)$. 
In the zeroth order approximation one can find this energy using only the first term in Eq. (\ref{eq:AbsRes1}). Under the conditions specified by  Eq. (\ref{eq:conditions}) one can expand all exponents in this term as $e^{x}\approx 1+x$ and then set the derivative of the expression with respect to energy $E$ to be equal to zero.  Then  we get $\sum_{i=1}^{n}(E^{(0)}-\phi_{i})/W_{i}^{3}=0$. This yields 
\begin{equation}
E^{(0)}=<\phi>_{W},
\label{eq:zerordmax}
\end{equation}
where $<A>_{W}=\left(\sum_{i=1}^{n}A_{i}/W_{i}^{3}\right)/\sum_{i=1}^{n}1/W_{i}^{3}$. 

One can approximate the shift of the maximum the due to the second term in Eq. (\ref{eq:AbsRes1}) as $E^{(1)} \approx – [d\beta^{(1)}(E^{(0)})/dE]/[d^{2}\beta^{(0)}(E^{(0)})/dE^{2}]$, where $\beta^{(0,1)}$ represents the  first (or second) terms in Eq. (\ref{eq:AbsRes1}). Both first and second derivatives can be evaluated at $E^{(0)}/W\approx 0$. They can be evaluated as  $d\beta^{(1)}(E^{(0)})/dE\approx (2/\sqrt{2\pi})\sum_{i\neq j}\tilde{V} _{ij}/(W_{i}^{2}W_{j})$ where $ \tilde{V} _{ij}=V_{ij}\cos(\theta_{i,j})/(1+\kappa_{ij})$
and  $d^{2}\beta^{(0)}(E^{(0)})/dE^{2}=(1/\sqrt{2\pi})\sum_{i=1}^{n}1/W_{i}^{3}$. Accordingly the position of the maximum $E_{max}$ of the function $\beta(E)$ can be expressed as 
\begin{equation}
E_{max}=<\phi>_{W}+ \frac{2\sum_{i\neq j}\tilde{V} _{ij}/(W_{i}^{2}W_{j})}{\sum_{i=1}^{n}1/W_{i}^{3}}.
\label{eq:Answer1}
\end{equation}
This result does not depend on the absolute value of the inhomogeneous broadening. 

In the simplified approximation one can consider only nearest neighbor interactions $V_{0}$.  Thus we assume $W_{1}=W_{n}=W_{ex}$, $W_{i}=W_{int}$, $i=2, 3, ... n-2$ and 
$\phi_{1}=\phi_{n}=\phi_{0}+\Delta$, $\phi_{i}=\phi_{0}$, $i=2, 3, ... n-1$. Then Eq. (\ref{eq:Answer1}) can be rewritten as 
\begin{eqnarray}
E_{max}=\phi_{0}+4\tilde{V}_{0}+
\nonumber\\
\frac{\frac{2\Delta}{W_{ex}^{3}}-4\tilde{V}_{0}\left(\frac{1}{W_{in}^{3}}+\frac{2}{W_{ex}^{3}}-\frac{1}{W_{i}W_{ex}^2}-\frac{1}{W_{i}^{2}W_{ex}}\right)}{\frac{n-2}{W_{in}^{3}}+\frac{2}{W_{ex}^{3}}}.
\label{eq:Answer2}
\end{eqnarray}
where $\tilde{V}_{0}=\frac{V_{0}\cos(\theta)}{1+\kappa_{0}}$,  $\kappa_{0}$ is the correlation coefficient for adjacent monomers. This result is different from the maximum shift obtained using  the convolution by  Gaussians method \cite{Santoro} because the latter method ignores the relative energy fluctuations of monomers. 

{\bf 4}. 
It is interesting to compare our predictions with experimental data. Here we consider the measurements of light absorption by poly-A - poly-T DNA hairpins endcapped by stilbene \cite{Fred} and containing $n$ $AT$ base pairs. To analyze the data we extracted positions of spectral maxima in $\beta(E)$ for $n=1$, $2$, $3$, $4$, $5$, $6$, $8$, $12$. We used the second maximum from the low energy side because the first maximum is due to stilbene absorption \cite{Fred,ab1}.  

Since all hairpins contained stilbene groups which also contributed to absorption the definition of maximum gets more accurate at large $n$. Therefore we fit experimental data using Eq. (\ref{eq:Answer2}) in the form $E_{max}=E_{\infty} - a/(bn+c)$ for $n=k, k+1, ...12$ choosing $k=1, 2, 3, ..$. The results of the data fit become almost insensitive to our choice of $k$ for $k\geq 3$ so we choose $k=3$ as a lower constraint. For smaller $k$ fitting parameters can change by $50\%$.  For any $k$ the best data fit is obtained setting $c=0$ ($W_{ex}=W_{in}$) in Eq. (\ref{eq:Answer2}) which suggests that the fluctuations of base pair energies are identical for DNA bases located at hairpin edges and inside the sequence. 
The reasonably good data fit is obtained setting (see Fig. \ref{fig:1})
\begin{equation}
E_{max}(n)=4.5242 - 0.135/n.
\label{eq:Ans3}
\end{equation}

One can assume that the energy difference $\Delta$ (Eq. (\ref{eq:Answer2})) of end and middle  hairpin $AT$ sites is negligible, e. g. due to pronounced screening effect of the static interactions leading to the ``diagonal'' energy difference $\Delta$. For dynamic ``off-diagonal'' interaction $V_{0}$ the screening is much smaller because it has to be taken at high exciton frequency $\omega \geq 10^{15}$s$^{-1}$. Then Eq. (\ref{eq:Answer2}) can be rewritten as  (cf. Eq. (\ref{eq:ANS0}))
\begin{eqnarray}
E_{max}=\phi_{0}+ \frac{4V_{0}\cos(\theta)}{1+\kappa_{0}} -\frac{4V_{0}\cos(\theta)}{n(1+\kappa_{0})}.
\label{eq:Answer5}
\end{eqnarray}	
Here we have assumed  $\theta=36^{0}$. Then the agreement of theory and experiment requires Eq. (\ref{eq:Ans3})
\begin{eqnarray}
V_{0}/(1+\kappa_{0}) \approx 0.0417 {\rm eV}.
\label{eq:Answer6}
\end{eqnarray}	
To verify this expectation we calculated the exciton coupling strength $V_{0}$ using the semiempirical program ZINDO (part of the Gaussian 03 package \cite{Gaussian}). The input geometry was obtained using $3$DNA software \cite{3DNA} with several $AT$ base pairs  placed at the neighboring distance $3.4 \AA$ with the twisting angle $36^{0}$. The ground state input geometry for each $AT$ pair has been optimized using density functional theory calculations based on B3LYP/6-31G** .  We studied the first ``bright'' excited  state of the $AT$ pair \cite{comment}. The energy of this state is found to be $4.2$eV. This energy is smaller than the experimental peak in the absorption spectrum at $4.5$eV by $0.3$eV which is a typical error for ZINDO \cite{Mark1}. 
The exciton coupling $V_{0}$ has been defined as half of the splitting of this bright excited energy level for two $AT$ pairs similarly to Ref. \cite{Newton}. The extracted value $V_{0} \approx 0.042$eV is very close to the experimental prediction Eq. (\ref{eq:Answer6}) if we ignore correlations between the excitation energies of adjacent base pairs ($\kappa_{0} \ll 1$). Since these correlations are due to the long-range electrostatic interactions, they should be weaker for neutral excitations than for charges where $\kappa_{0}\approx 1/2$ \cite{Voityuk}. Our results agree qualitatively with the analysis of excitons in the single strand DNA made using $A$ bases that were reported in Refs. \cite{alternative,Santoro}. Excitonic coupling has been estimated as $0.053$eV in that system.  

It is important that the exciton coupling $V_{0}$ is positive leading to an increase of the absorption maximum energy $E_{max}$ with increasing number of monomers, i. e. a blue shift of the absorption spectrum in agreement with the experiment \cite{Fred,DM}. This is because of the dipolar nature of exciton coupling. The oscillator strength is larger for quantum states with larger transition dipole moments. These states  have higher energy because the interaction of parallel $AT$ pair dipole moments is repulsive in DNA geometry. The red shift obtained in Ref. \cite{George} is probably the consequence of self-interaction errors in the TDDFT calculations that were used.


{\bf 5}. 
We have investigated the inhomogeneously broadened absorption spectrum of a molecule made of $n$ identical monomers. The shift of the maximum of the rescaled absorption $\beta(E)=\alpha(E)/E$ with $n$ is determined entirely by the average monomer energies and their exciton couplings. It is insensitive to the inhomogeneous broadening and can be expressed by a simple algebraic form Eq. (\ref{eq:Answer2}). Thus exciton coupling can be characterized using inhomogeneously broadened absorption spectra. 

A related but challenging problem is whether the exciton quantum state is localized within a single $AT$ base pair or it is extended through several base pairs \cite{DM,George}. This localization is defined by the ratio of exciton coupling strength $V_{0}$ and inhomogeneous broadening $W$. If $V_{0}/W \ll 1$ this should be the regime of strong localization \cite{ab2}. We can use our estimate $V_{0}\sim 0.04$eV and compare it to the width of the DNA absorption maximum $W \sim 0.2$eV \cite{Fred}.   Then $V_{0}<W$ so the exciton wavefunction is probably localized within almost a single site. Note that a recent study of this effect for single stranded DNA considered the behavior of a large number of eigenstates, and came to the opposite conclusion; however this is not necessarily in contradiction with the present analysis, which only considered a single excited state \cite{DM,George}. More excited states need to be considered to verify our expectations.


This work is supported by the NSF CRC Program, Grant No. 0628092. JD and CH also acknowledge the support of the Lurcy Fund, and CH acknowledges the support of the Newcomb and Provost Funds of Tulane University. 
The authors are grateful to Fred Lewis and John Perdew for useful stimulating discussions and Scott Knowles, Balamurugan Desinghu, Michael Armbruster and Amy Finch for contribution to our computations.



\begin{references}
\bibitem{Fred}
F. D. Lewis, L. G.  Zhang, X. Y. Liu, X. B. Zuo, D. M. Tiede, 
H. Long, G. C.  Schatz,  J. Am. Chem. Soc.  {\bf 127}, 14445 (2005).



\bibitem{books}
I. Tinoco, Jr., J. Am. Chem. Soc.,
1960, 82, 4785–90; B. Bouvier, J. P. Dognon, R. Lavery, D. Markovitsi, P.Milli´e, D. Onidas, K. Zakrzewska, J. Phys. Chem. B {\bf 107}, 13512 (2003). 



\bibitem{ab1}
D. Balamurugan, F. D. Lewis, A. L. Burin, J. Phys. Chem. {\bf 111}, 3982 (2007).  

\bibitem{ab2}
A. L. Burin, D. B. Uskov, to appear in J. Chem. Phys. (2008).

\bibitem{LL3}
L. D. Landau, E. M. Lifshitz, Quantum Mechanics: non-relativistic theory, 	 London, Pergamon Press, 1958.

\bibitem{Santoro}
F. Santoro, V. Barone, R. Improta, PNAS {\bf 104}, 9931 (2007). 

\bibitem{3DNA}
W. K. Olson et al., J. Mol. Biol. {\bf 313}(1), 229 (2001). 

\bibitem{Gaussian}
Gaussian 03, Revision C.02, M. J. Frisch, et al, Gaussian, Inc., Wallingford CT, 2004.

\bibitem{comment} This is the third excited singlet state. According to both ZINDO and TDDFT calculations (see e.g. D. Varsano, et al,  J. Phys. Chem. B {\bf 110}, 7129 (2006) and references therein) first few excited states are dark, i. e. their oscillator strengths are  smaller than $10^{-3}$. The analysis of excited states in detail will be reported separately.  

\bibitem{Mark1}
G. R. Hutchison, M. A. Ratner, T. J. Marks, J. Phys. Chem. A, {\bf 106}, 10596 (2002). 

\bibitem{Newton}
R. J. Cave, M. D. Newton, Chem. Phys. Lett. {\bf 249}, 15 (1996).





\bibitem{Voityuk}  A. A. Voityuk, J. Chem. Phys. {\bf 122}, 204904  (2005). 

\bibitem{alternative}
Hu, L. H.; Zhao, Y.; Wang, F.; Chen, G. H.; Ma, C.; Kwok, W.-M.;
Phillips, D. L., J. Phys. Chem. B {\bf 111}, 11812 (2007).


\bibitem{DM} D.Markovitsi, D. Onidas, T. Gustavsson, F. Talbot and E. Lazzarotto,
J. Am. Chem. Soc. {\bf 127}, 17130 (2005).

\bibitem{George} S. Tonzani, G. C. Schatz, J. Amer. Chem. Soc. {\bf 130}, 7607 (2008).

\end{references}
\end{document}